\documentclass[a4paper, 11pt]{article}

\usepackage{a4wide}
\usepackage{amsmath}
\usepackage{amssymb}
\usepackage{color}
\usepackage{epsfig}
\usepackage{float}
\usepackage{graphicx}

\begin{document}

\baselineskip = 15pt
\parskip 1.8mm
\renewcommand{\baselinestretch}{1.0}


\newcommand{\Ga}{{\Gamma}}
\newcommand{\De}{{\Delta}}
\newcommand{\Lm}{{\Lambda}}
\newcommand{\Om}{{\Omega}}

\newcommand{\al}{{\alpha}}
\newcommand{\bt}{{\beta}}
\newcommand{\ga}{{\gamma}}
\newcommand{\de}{{\delta}}
\newcommand{\ep}{{\epsilon}}
\newcommand{\vep}{{\varepsilon}}
\newcommand{\zt}{{\zeta}}
\newcommand{\te}{{\theta}}
\newcommand{\ka}{{\kappa}}
\newcommand{\lm}{{\lambda}}
\newcommand{\vpi}{{\varpi}}
\newcommand{\sig}{{\sigma}}
\newcommand{\vphi}{{\varphi}}
\newcommand{\om}{{\omega}}

\newcommand{\alt}{{\rm alt}}
\newcommand{\bdy}{{\rm bdy}}
\newcommand{\bsa}{{\boldsymbol{a}}}
\newcommand{\bsb}{{\boldsymbol{b}}}
\newcommand{\bsD}{{\boldsymbol{D}}}
\newcommand{\bsk}{{\boldsymbol{k}}}
\newcommand{\bsM}{{\boldsymbol{M}}}
\newcommand{\bulk}{{\rm bulk}}
\newcommand{\cN}{{\mathcal{N}}}
\newcommand{\cont}{{\rm cont.}}
\newcommand{\ct}{{\rm ct}}
\newcommand{\cdN}{{\mathcal{N}_d}}
\newcommand{\dg}{{\dagger}}
\newcommand{\I}{{\rm I}}
\newcommand{\IR}{{\rm \, IR}}
\newcommand{\lan}{{\langle}}
\newcommand{\nab}{{\nabla}}
\newcommand{\nn}{{\nonumber}}
\newcommand{\ol}{\overline}
\newcommand{\pd}{{\partial}}
\newcommand{\pr}{{\prime}}
\newcommand{\prl}{{\parallel}}
\newcommand{\R}{{\rm R}}
\newcommand{\ra}{{\rm a}}
\newcommand{\rad}{{\rm rad}}
\newcommand{\ran}{{\rangle}}
\newcommand{\rs}{{\rm s}}
\newcommand{\Slash}[1]{{\ooalign{\hfil/\hfil\crcr$#1$}}} 
\newcommand{\srel}[2]{{\stackrel{\scriptstyle #1}{\scriptstyle #2}}}
\newcommand{\std}{{\rm std}}
\newcommand{\U}{{\rm U}} 
\newcommand{\ul}{\underline}
\newcommand{\UV}{{\rm UV}}
\newcommand{\wg}{{\wedge}}
\newcommand{\wh}{\widehat}
\newcommand{\wt}{\widetilde}

\def\Log{\mathop{\rm Log}}
\def\Spin{\mathop{\rm Spin}}
\def\SO{\mathop{\rm SO}}
\def\O{\mathop{\rm O}}
\def\SU{\mathop{\rm SU}}
\def\U{\mathop{\rm U}}
\def\Sp{\mathop{\rm Sp}}
\def\SL{\mathop{\rm SL}}
\def\GL{\mathop{\rm GL}}

\def\det{\mathop{\rm det}\nolimits}
\def\sign{\mathop{\rm sign}\nolimits}
\def\mod{\mathop{\rm mod}\nolimits}
\def\tr{\mathop{\rm tr}\nolimits}
\def\diag{\mathop{\rm diag}\nolimits}
\def\Re{\mathop{\rm Re}\nolimits}
\def\Im{\mathop{\rm Im}\nolimits}
\def\Tr{\mathop{\rm Tr}\nolimits}
\def\bbra{{\langle\kern-2.5pt\langle}}
\def\kket{{\rangle\kern-2.5pt\rangle}}
\def\Bbra{{\Big\langle\kern-3.5pt\Big\langle}}
\def\Kket{{\Big\rangle\kern-3.5pt\Big\rangle}}

\newcommand{\ed}{{\rm d}}

\thispagestyle{empty}
\addtocounter{page}{-1}
\vskip-0.35cm
\vspace*{0.2cm} 
\begin{center}
{\LARGE \bf Note on the self-duality of gauge fields in topologically nontrivial spacetime
}
\end{center}

\vspace*{0.1cm}

\vspace*{1.0cm} 
\centerline{\large\bf 
Hiroshi~Isono}
\vspace*{0.7cm}
\centerline{\it Department of Physics, National Tsing Hua University, Hsinchu, TAIWAN}
\vspace*{0.4cm}
\centerline{\tt email: hiroshi.isono81@gmail.com
}

\vspace*{0.8cm}
\centerline{\bf Abstract}
\vspace*{0.3cm}
\vspace*{0.5cm} 

We show the derivation of the self-duality relation of abelian higher-form gauge field strength in the topologically nontrivial spacetime background. The so-called Pasti-Sorokin-Tonin action for the self-dual abelian gauge field assumes that the spacetime topology is trivial to derive the self-duality relation using a gauge transformation of the action. In this paper we find a new gauge transformation of the same theory and show that this new gauge transformation enables us to derive the self-duality relation even in the topologically nontrivial spacetime background.

\vspace{3ex}


\section{Introduction}
This paper considers 
the Lorentz covariant action, so-called the PST (Pasti-Sorokin-Tonin) action 
\cite{Pasti:1996vs,Pasti:1997gx}, for the self-dual field strength of abelian higher-form gauge field on a general spacetime manifold. Especially we focus on the derivation of the self-duality relation in the case where the spacetime has nontrivial cycles.

Recently \cite{Mori:2014tca} the PST action for single M5-brane was applied to the check of AdS$_7$/CFT$_6$ correspondence for
the Wilson surface operators of 6 dimensional (2,0) $A_n$ superconformal field theory (or its $S^1$-reduction to 5 dimensional maximally supersymmetric Yang-Mills theory). This work obtained the nice correspondence, but there was a small subtlety. We explain this briefly.

In the gravity side, they computed the on-shell value of the PST action of a probe M5-brane which wraps $S^3$
in the AdS$_7\times S^4$ background. Thus the worldvolume of the probe M5-brane has a nontrivial 3-cycle.

On the other hand, the recipe to derive the self-duality relation in the PST formalism is to find a general solution to the equation of motion and to gauge-transform the solution to produce the self-duality relation, the technical detail of which will be given in the next section.
For this recipe to work, we have to assume that the worldvolume is topologically trivial
so that the Poincar\'{e}'s lemma holds, otherwise the gauging-away to produce the self-duality relation does not work.
In other words, the self-duality relation cannot be derived when the worldvolume has nontrivial cycles.

In summary, \cite{Mori:2014tca} obtained the nice correspondence using the PST action in the case where the self-duality relation cannot be derived,
though the M5-brane worldvolume theory must have the self-dual gauge field.
This situation may motivate us to think that the self-duality can be derived even when the worldvolume has nontrivial cycles. 

In this paper we claim that the self-duality relation can be derived even when the worldvolume has nontrivial cycles.
To show this we find a new gauge transformation \eqref{new}, which may be regarded as a modification of the gauge transformation which is used in the conventional derivation of the self-duality relation.\footnote{A similar result was obtained in \cite{Bekaert:1998yp} for not manifestly Lorentz covariant action.}
In the next section we will show the details of the new gauge transformation and the derivation of the self-duality relation.

\section{PST action, gauge transformations and the self-duality}
\paragraph{PST action}
We consider the Lorentz covariant action, so-called the PST action, for the self-dual field strength of abelian $2n$-form gauge field on a general manifold of dimensionality $D=4n+2~(n=0,1,2)$ with metric $g$.
Properties of the action and the gauge transformations will be shown 
in an index-free manner. For definitions and notations, see the appendix.

Let $M$ be a $(2n+1)$-form field strength. 
In this paper we consider the case $M=dA$ for simplicity,  where $A$ is the $2n$-form gauge potential.\footnote{
When applied to the 10D IIB supergravity, the field strength is given by 
\cite{Dall'Agata:1997ju,Dall'Agata:1998va}
\begin{align}
M:=dA+(B\wg H'-B'\wg H)/2, \quad\quad H:=dB, \quad H':=dB', \nn
\end{align}
where $(B,B')$ is an $\SL(2,\mathbb{Z})$-doublet.
Then, for the gauge invariance which we will explain shortly, we need to add a topological term $A\wg H\wg H'$. This term itself had already been known before the PST action for the IIB SUGRA appeared 
\cite{Dall'Agata:1997ju,Dall'Agata:1998va}. 
Our new gauge invariance under 
\eqref{new} still holds in this case.
}
The self-duality relation is $M=\ast M$.
The PST action is given by \cite{Pasti:1996vs,Pasti:1997gx,Pasti:2012wv}
\begin{align}
S=\int \mathcal{L} = \int e_{\wh{v}}\iota_v(M-\ast M)\wg M,
\label{pst}
\end{align}
where we introduced a 1-form field $\wh{v}$ with an auxiliary 0-form field $a$,
and its dual vector field $v$ obtained by raising the index of $\wh{v}$,
\begin{align}
\wh{v} &:= \frac{da}{\sqrt{(\pd a)^2}}, \quad\quad
v := \frac{\pd^aa}{\sqrt{(\pd a)^2}}\pd_a, \quad\quad
(\pd a)^2 := g^{ab}\pd_aa\pd_ba. \nn
\end{align}
For the action in components, see the appendix.
The action \eqref{pst} has three gauge symmetries:
\begin{align}
&{} \de_1a=0, \quad\quad \de_1A=d\Lm, \label{g1} \\
&{} \de_2a=0, \quad\quad \de_2A=da\wg\Phi, \label{g2} \\
&{} \de_3a=\vphi, \quad\quad \de_3A=\frac{\vphi}{\sqrt{(\pd a)^2}}\iota_v(M-\ast M), \label{g3}
\end{align}
where $\Lm$ and $\Phi$ are $(2n-1)$-forms and $\vphi$ is a 0-form.
The first gauge transformation $\de_1$ is the usual one.
The second gauge symmetry $\de_2$ will be used to
derive the self-duality condition from the equation of motion, as will be shown later.
The third gauge symmetry $\de_3$ is used to eliminate the auxiliary field $a$.
This is the origin of the modified Lorentz invariance of the actions \cite{Perry:1996mk,Chen:2010jgb}
of self-dual gauge fields in which the Lorentz symmetry is not manifest.

The variation of the action \eqref{pst} reads
\begin{align}
\de\mathcal{L} 
&= 
\frac{1}{\sqrt{(\pd a)^2}}
d\de a\wg \wh{v}\wg
\left[ \iota_v(M-\ast M)\wg \iota_v(M-\ast M) \right]
-2\de M\wg\wh{v}\wg\iota_v(M-\ast M)
-M\wg\de M
\nn\\
&= 
-\de a \cdot d\left[
\frac{1}{\sqrt{(\pd a)^2}}
\wh{v}\wg\iota_v(M-\ast M)\wg \iota_v(M-\ast M) \right]
+2\de A\wg d\left[\wh{v}\wg\iota_v(M-\ast M)\right] \nn\\
&{} \quad
+(\mbox{ total derivatives }),
\label{totvar}
\end{align}
where we used the formulae \eqref{fml} and the variations of $\wh{v}$ and $\iota_v$ on any $n$-form 
\begin{align}
\de\wh{v} &= \frac{1}{\sqrt{(\pd a)^2}}\iota_ve_{\wh{v}}d\de a, \nn\\
\de(\iota_v\om_n)
&= \frac{1}{\sqrt{(\pd a)^2}}(-)^n\ast(\iota_ve_{\wh{v}}d\de a\wg\ast^{-1}\om_n)+\iota_v\de\om_n. \nn
\end{align}
It is very easy to see the invariance under the first gauge transformation $\de_1$ since $\de_1M=0$.
Let us consider the second gauge transformation $\de_2$.
The $\de a$-term of \eqref{totvar} vanishes under $\de_2$.
We can show that $\de M$-term also vanishes
since $\de M$ is a product of $da$ and a $2n$-form and $\de M$-term contains $\wh{v}$. 
Then the action is invariant under $\de_2$.
Finally we can show the invariance under the third gauge transformation $\de_3$
using the relation
\begin{align}
d\left(\frac{1}{\sqrt{(\pd a)^2}}\right)\wg\wh{v}
= \frac{1}{\sqrt{(\pd a)^2}}d\wh{v}. \nn
\end{align}
\paragraph{derivation of self-duality relation}
The equations of motion can be read from \eqref{totvar},
\begin{align}
\de a \quad : &{} \quad 
d\left[ \frac{1}{\sqrt{(\pd a)^2}} \wh{v}\wg\iota_v(M-\ast M)\wg \iota_v(M-\ast M) \right] = 0, \label{eoma} \\
\de A \quad : &{} \quad d\left[ \wh{v}\wg\iota_v(M-\ast M) \right]=0. \label{eomA}
\end{align}
Note that \eqref{eoma} holds identically once \eqref{eomA} holds.
Therefore the independent equation of motion is \eqref{eomA} only.
The general solution to this takes the form 
\begin{align}
\wh{v}\wg\iota_v(M-\ast M)=da\wg(d\eta+\om), \label{solA}
\end{align}
where $\om$ is closed but not exact,
since, using \eqref{eomA}, 
\begin{align}
\wh{v}\wg\iota_v(M-\ast M)=da\wg(2n\mbox{-form})
\rightarrow
d(2n\mbox{-form})=0
\rightarrow
(2n\mbox{-form})=d\eta+\om. \nn
\end{align}
As can be seen, the $\om$-term appears when the spacetime has nontrivial cycles.
On the other hand, the $\de_2$-variation of \eqref{solA} reads 
$\de_2[\wh{v}\wg\iota_v(M-\ast M)]=da\wg d\Phi$.
Then $\de_2$ allows us to eliminate the $d\eta$-part of the solution \eqref{solA} by setting $\Phi=-\eta$,
while the $\om$-part cannot be eliminated by $\de_2$.
Therefore the self-duality relation cannot be derived when the spacetime background has nontrivial cycles.
\paragraph{new gauge transformation}
Actually we can show that the action \eqref{pst} is invariant under
\begin{align}
\de_2' a=0, \quad\quad \de_2' A=a\xi, \label{new}
\end{align}
where $\xi$ is closed but not necessarily exact $2n$-form.
Then $\de_2' M=da\wg\xi$.
Since this has $da$, the variation of the action $\eqref{totvar}$ vanishes.
Thus this transformation can also be regarded as a gauge symmetry.
Now the $\de_2'$-variation of \eqref{solA} reads $\de_2'[\wh{v}\wg\iota_v(M-\ast M)]=-da\wg\xi$.
Therefore we can eliminate the $\om$-part of the solution.
Since $\de_2'$ can also be used to eliminate the $d\eta$-part of the solution \eqref{solA}
when $\xi$ is an exact form,
the new transformation $\de_2'$ plays the same role as $\de_2$.
Thus we can adopt $\de_2'$ instead of $\de_2$.
Then the solution \eqref{solA} can be made into 
\[
e_{\wh{v}}\iota_v(M-\ast M)=0.
\]
Inserting $\ast\ast=1$ into this and using $e_{\wh{v}}\iota_v\ast=\ast\iota_ve_{\wh{v}}$,
which can be derived from \eqref{fml}, we obtain
\[
\iota_ve_{\wh{v}}(M-\ast M)=0.
\]
Then summing the last two equations and using $\iota_{v}e_{\wh{v}}+e_{\wh{v}}\iota_v=1$, 
we get the self-duality relation
\[
M=\ast M.
\]
Therefore we can obtain the self-duality relation even when the spacetime has nontrivial cycles.

\section{Conclusion and Discussion}
In this paper we found the new gauge transformation which allows us to derive the self-duality relation even when the spacetime has nontrivial cycles. 
Our result can be applied to any actions and equations of motion for self-dual gauge fields which use PST-like mechanism to derive the self-duality relations as the solution to the equations of motion, including 
DBI-like extensions such as \cite{Pasti:1996vs,Pasti:1997gx,Pasti:2012wv},
actions with more than one auxiliary fields such as \cite{Pasti:2009xc},
actions which do not have manifest Lorentz covariance such as \cite{Perry:1996mk,Chen:2010jgb,Henneaux:1988gg},
the formalisms for non-abelian self-dual gauge fields such as \cite{Chu:2012um}.

Our result will enlarge the range of applications of the PST action to physical systems in various spacetime backgrounds. Especially it would be interesting to apply the formalism to the AdS/CFT correspondence with 10 dimensional IIB supergravity with self-dual Ramond-Ramond 5-form field strength, which we are now studying \cite{isono}.

It will be interesting to investigate the relation between the cohomology of the spacetime manifold and the structure of the physical states of systems of the self-dual gauge fields.
The 2-dimensional case on a torus has been studied in \cite{Chen:2013gca}, where
the BRST analysis of the Floreanini-Jackiw action \cite{Floreanini:1987as} on a torus was carried out and 
it was shown that the self-duality relation holds on the physical states.
Our result can be regarded as the classical proof of this in more general spacetime backgrounds.
But the quantum analysis with the explicit BRST charge and the state space is much more important and strong.
Especially it would be interesting to see whether there are any nontrivial relations between the spacetime cohomology and the BRST cohomology for self-dual gauge field theories by the quantum BRST analysis.

\paragraph{Note added:}
We have already found the new result in this paper long before,
and were preparing a paper \cite{isono} which includes not only the result in this paper 
but also a new physical application of the PST action.
Then we found the paper \cite{Bandos:2014bva}, which proposes a gauge transformation, which is similar to our new gauge transformation.

\section*{Acknowledgement} 
We would like to thank Chong-Sun Chu, Sheng-Lan Ko, Hironori Mori, Takeshi Morita, and Satoshi Yamaguchi 
for fruitful discussions.

\appendix

\section{Notations}
Here we give the definitions of symbols used in the main part.
We fix the dimensionality of the manifold to be $D$ and the metric to be $g$.
A $k$-form $\om_k$ is related to its components by
\[
\om_k=\frac{1}{k!}\om_{a_1\cdots a_k}dx^{a_1}\wg\cdots\wg dx^{a_k}.
\]
The Hodge star, which transforms $k$-form to $(D-k)$-form, is defined so that
\begin{align}
\eta_k\wg\ast\te_k 
&= \frac{1}{k!}\eta_{a_1\cdots a_k}\te^{a_1\cdots a_k}\sqrt{|g|}dx^1\wg\cdots\wg dx^D.
\end{align}
The Hodge star satisfies $\ast\ast\om_k=(-)^{k(D-k)}\sign(g)\om_k$.

The exterior product $e_{\te}$ with a 1-form $\te$ maps $k$-form to $(k+1)$-form
and is defined by
\begin{align}
e_{\te}\om &:= \te\wg\om.
\end{align}
The interior product $\iota_v$ with a vector $v=v^a\pd_a$ maps $k$-form to $(k-1)$-form
and is defined by
\begin{align}
\iota_v\om_k &:= \frac{1}{(k-1)!}v^a\om_{aa_1\cdots a_{k-1}}dx^{a_1}\wg\cdots\wg dx^{a_{k-1}}.
\end{align}
We will use the following useful formulae in deriving the variation of the action:
on any $k$-form 
\begin{align}
\iota_v=(-)^{k+1}\ast^{-1}e_{\wh{v}}\ast=(-)^{D+k}\ast e_{\wh{v}}\ast^{-1}, \quad\quad 
e_{\wh{v}}=(-)^k\ast^{-1}\iota_v\ast=(-)^{D+k+1}\ast\iota_v\ast^{-1}.
\label{fml}
\end{align}
The action \eqref{pst} in components reads
\begin{align}
S=\int\!d^Dx\frac{\sqrt{-g}}{(\pd a)^2}\pd^aa(M-\ast M)_{aa_1\cdots a_{2n}}(\ast M)^{a_1\cdots a_{2n}b}\pd_ba.
\end{align}

\end{document}